\documentclass[twocolumn,aps,prd,superscriptaddress,nofootinbib,floatfix]{revtex4-2}

\usepackage{graphicx}
\usepackage{multirow}
\usepackage[colorlinks=true,citecolor=blue,urlcolor=blue,linkcolor=blue]{hyperref}

\usepackage{amssymb}
\usepackage{amsmath}
\usepackage{xcolor}

\newcommand{\nn}{\nonumber}
\newcommand{\beq}{\begin{equation}}
\newcommand{\eeq}{\end{equation}}
\newcommand{\beqa}{\begin{eqnarray}}
\newcommand{\eeqa}{\end{eqnarray}}

\newcommand{\GeV}{{\rm GeV}}

\def\ov{\overline}
\def\lqcd{\Lambda_{\rm QCD}}
\def\d{{\rm d}}

\newcommand{\Bbar}{\,\overline{\!B}{}}
\newcommand{\Dbar}{\,\overline{\!D}{}}
\newcommand{\Kbar}{\,\overline{\!K}{}}
\def\B0bar{\Bbar{}^0}
\def\D0bar{\Dbar{}^0}
\def\K0bar{\Kbar{}^0}
\newcommand{\Lb}{\Lambda_b}
\newcommand{\Lc}{\Lambda_c}

\def\OMIT#1{{}}
\makeatletter
\g@addto@macro\bfseries{\boldmath}
\let\Hy@backout\@gobble
\makeatother

\begin{document}

\preprint{CALT-TH-2022-023}

\title{Interpreting LHCb's \texorpdfstring{$\Lambda_b\to \Lambda_c\tau\bar\nu$}{Lambda-b -> Lambda-c tau nu} 
measurement and puzzles in semileptonic \texorpdfstring{$\Lambda_b$}{Lambda-b} decays}

\author{Florian U.\ Bernlochner}
\affiliation{Physikalisches Institut der Rheinischen Friedrich-Wilhelms-Universit\"at Bonn, 53115 Bonn, Germany}

\author{Zoltan Ligeti}
\affiliation{Ernest Orlando Lawrence Berkeley National Laboratory, 
University of California, Berkeley, CA 94720, USA}
\affiliation{Berkeley Center for Theoretical Physics, 
Department of Physics,
University of California, Berkeley, CA 94720, USA}

\author{Michele Papucci}
\affiliation{Burke Institute for Theoretical Physics, 
California Institute of Technology, Pasadena, CA 91125, USA}

\author{Dean J.\ Robinson}
\affiliation{Ernest Orlando Lawrence Berkeley National Laboratory, 
University of California, Berkeley, CA 94720, USA}
\affiliation{Berkeley Center for Theoretical Physics, 
Department of Physics,
University of California, Berkeley, CA 94720, USA}

\begin{abstract}

Normalizing the recent LHCb measurement of $\Lambda_b \to \Lambda_c \tau \bar\nu$ to 
the standard model (SM) prediction for the $\Lambda_b \to \Lambda_c \mu \bar\nu$ rate, instead of a LEP measurement, provides a more consistent comparison with the SM prediction for the lepton flavor universality ratio $R(\Lambda_c)$. 
This modestly increases $R(\Lambda_c)$ compared to the quoted LHCb result, such that it no longer hints at a suppression compared to the SM,
which would be hard to accommodate in new physics scenarios that enhance $R(D^{(*)})$.
We point out that the fraction of excited states in inclusive semileptonic $\Lambda_b$ decay may be significantly greater than the corresponding fraction in $B$ decays. Possible implications are speculated upon.

\end{abstract}

\maketitle

The LHCb Collaboration recently observed for the first time the
$\Lambda_b \to \Lambda_c \tau \bar\nu$ decay at $6.1\sigma$ significance~\cite{LHCb:2022piu}. 
This is an impressive and particularly timely measurement, 
because of persistent hints~\cite{HFLAV:2019otj} of lepton flavor universality violation (LFUV) 
in the ratios $R(H_c) = \Gamma(H_b \to H_c\tau\bar\nu)/\Gamma(H_b \to H_c \ell\bar\nu)$, $\ell=e,\mu$,
where $H_{b,c}$ denotes a beauty and charmed hadron, respectively. 
Measurements of $R(D^{(*)})$ have exceeded the SM expectations by about 20\%, and are in more than $3\sigma$ tension with the SM.

The $\Lb^0 \to \Lc^+ l \bar\nu$ baryon decay provides a complementary probe of the underlying physics, but subject to (partly) different uncertainties.  
LHCb has recently published~\cite{LHCb:2022piu}
\beq\label{LHCbresult}
R(\Lc) = 0.242 \pm 0.026 \pm 0.040 \pm 0.059\,.
\eeq
In order to assess agreement with the SM, the LHCb analysis compares Eq.~\eqref{LHCbresult} with the SM prediction~\cite{Bernlochner:2018kxh, Bernlochner:2018bfn}
\beq\label{RLcSM}
    R(\Lambda_c)_{\text{SM}} = 0.324 \pm 0.004\,.
\eeq
This prediction is obtained by fitting a heavy quark effective theory (HQET) based parametrization of the $\Lb \to \Lc$ form factors~\cite{Isgur:1990pm, Falk:1992ws} at ${\cal O}(\lqcd^2/m_c^2)$ to the $\d\Gamma(\Lb \to \Lc \mu \bar\nu)/\d q^2$ shape measurement~\cite{LHCb:2017vhq} plus lattice QCD (LQCD) predictions~\cite{Detmold:2015aaa}.

The LHCb result~\eqref{LHCbresult} is derived from the measurement
\beq
    {\cal K}(\Lc) 
     \equiv \frac{{\cal B}[\Lb \to \Lc \tau \bar\nu]}{{\cal B}[\Lb \to \Lc 3\pi]} 
     = 2.46 \pm 0.27 \pm 0.40\,,
\eeq
using $\Lb \to \Lc 3\pi$ as a normalization channel to obtain
\beq\label{LHCbtau}
{\cal B}(\Lambda_b \to \Lambda_c \tau \bar\nu) = (1.50 \pm 0.16 \pm 0.25 \pm 0.23)\,\%\,,
\eeq
in which the third uncertainty comes from the normalization mode.
The first and second uncertainties in Eq.~\eqref{LHCbresult} are statistical and systematic, respectively, 
while the third one incorporates the 
uncertainty in the branching fraction
${\cal B}(\Lb \to \Lc \mu \bar\nu) = (6.2 \pm 1.4)\,\%$~\cite{ParticleDataGroup:2020ssz},
which is used to determine $R(\Lc)$ from Eq.~\eqref{LHCbtau}.
This branching fraction is the result of a PDG fit~\cite{ParticleDataGroup:2020ssz}
incorporating a measurement by the DELPHI Collaboration~\cite{DELPHI:2003qft},
${\cal B}(\Lb \to \Lc \mu \bar\nu) = (5.0^{+1.1}_{-0.8}{} ^{+1.6}_{-1.2})\,\%$~\cite{DELPHI:2003qft},
together with correlated measurements for $\Lb \to \Lc \pi$, $\Lb \to \Lc K$, and $\Lb \to \Lc 3\pi$.
The DELPHI measurement is anticorrelated with the ${\cal B}(\Lc \to p K \pi)$ and the ${\cal B}(b \to \Lb)$ production fraction.
The values used for these were $(5.0 \pm 1.3)\%$ and $(10.8 \pm 2.0)\%$ respectively, versus the current world average ${\cal B}(\Lc \to p K \pi) = (6.28 \pm 0.32)\%$ and ${\cal B}(b \to \Lb) = (8.4 \pm 1.1)\%$~\cite{ParticleDataGroup:2020ssz}.
Though these two effects would approximately cancel in the ${\cal B}(\Lb \to \Lc \mu \bar\nu)$ central value,
they will also reduce the systematic uncertainties of the DELPHI measurement, and thus likely pull the PDG fit to lower values.
Moreover, within the PDG fit, the $\Lb \to \Lc \mu \bar\nu$ branching fraction is also mildly correlated to the $\Lb \to \Lc 3\pi$ normalization channel.
Thus, the result in Eq.~\eqref{LHCbresult} might not incorporate potentially important corrections or correlations.

We point out that it is more robust and precise to compare 
Eq.~\eqref{LHCbtau} with the SM prediction for $\Gamma(\Lb \to \Lc l \bar\nu)$, $l = \tau$ (or $\mu$ or $e$), which can be trivially derived from the very same fits in Ref.~\cite{Bernlochner:2018kxh} as used to obtain Eq.~\eqref{RLcSM},
and can also be computed using the \texttt{Hammer} library \cite{Bernlochner:2020tfi,bernlochner_florian_urs_2022_5828435}.
These predictions are more precise than the DELPHI measurement or the PDG fit,
even accounting for the uncertainty from the inclusive versus exclusive values for $|V_{cb}|^2$~\cite{HFLAV:2019otj,Bordone:2021oof}.
Explicitly, they are 
\begin{align}
\Gamma(\Lambda_b \to \Lambda_c \mu \bar\nu) &= |V_{cb}|^2\, (14.81 \pm0.69) \times 10^{-12}\, \GeV\,, \nn\\
\Gamma(\Lambda_b \to \Lambda_c \tau \bar\nu) &= |V_{cb}|^2\, (4.79 \pm 0.21) \times 10^{-12}\, \GeV\,.
\end{align}
The correlations between $\Gamma(\Lambda_b \to \Lambda_c \mu \bar\nu)/|V_{cb}|^2 : \Gamma(\Lambda_b \to \Lambda_c \tau \bar\nu)/|V_{cb}|^2 : R(\Lambda_c)$ are
\beq
\begin{pmatrix} 
~~1.~~  &  0.9723  &  ~-0.3332~ \cr
\text{---}  &  1.  &  -0.1038 \cr
\text{---}   &  \text{---}  &  1.
\end{pmatrix} ,
\eeq
and the predictions for the branching fractions are
\begin{subequations}
\label{BLRSpred}
\begin{align}
{\cal B}(\Lambda_b \to \Lambda_c \mu \bar\nu) &= |V_{cb}/0.04|^2\, (5.27\pm0.25)\, \% \,, \label{BLRSpredmu}\\
{\cal B}(\Lambda_b \to \Lambda_c \tau \bar\nu) &= |V_{cb}/0.04|^2\, (1.70\pm0.08)\, \% \,.\label{BLRSpredtau}
\end{align}
\end{subequations}
The latter prediction is in good agreement with the LHCb measurement in Eq.~\eqref{LHCbtau},
over the exclusive to inclusive range for $|V_{cb}|$.
That is, using Eqs.~\eqref{LHCbtau} and~\eqref{BLRSpredtau}, 
one finds $|V_{cb}| = (37.5 \pm 4.5) \times 10^{-3}$.
This is compatible with $|V_{cb}|$ determined both from exclusive and inclusive semileptonic decays, for which we use
\begin{align}\label{Vcbval}
|V_{cb}|_{\text{excl}} & = (39.10 \pm 0.50)\times 10^{-3}~\mbox{\cite{HFLAV:2019otj}}\,,\nn\\
|V_{cb}|_{\text{incl}} & = (42.16 \pm 0.51)\times 10^{-3}~\mbox{\cite{Bordone:2021oof}}\,.
\end{align}

\begin{figure}[t!]
	\includegraphics[width=0.95\columnwidth, clip, bb=15 45 585 390]{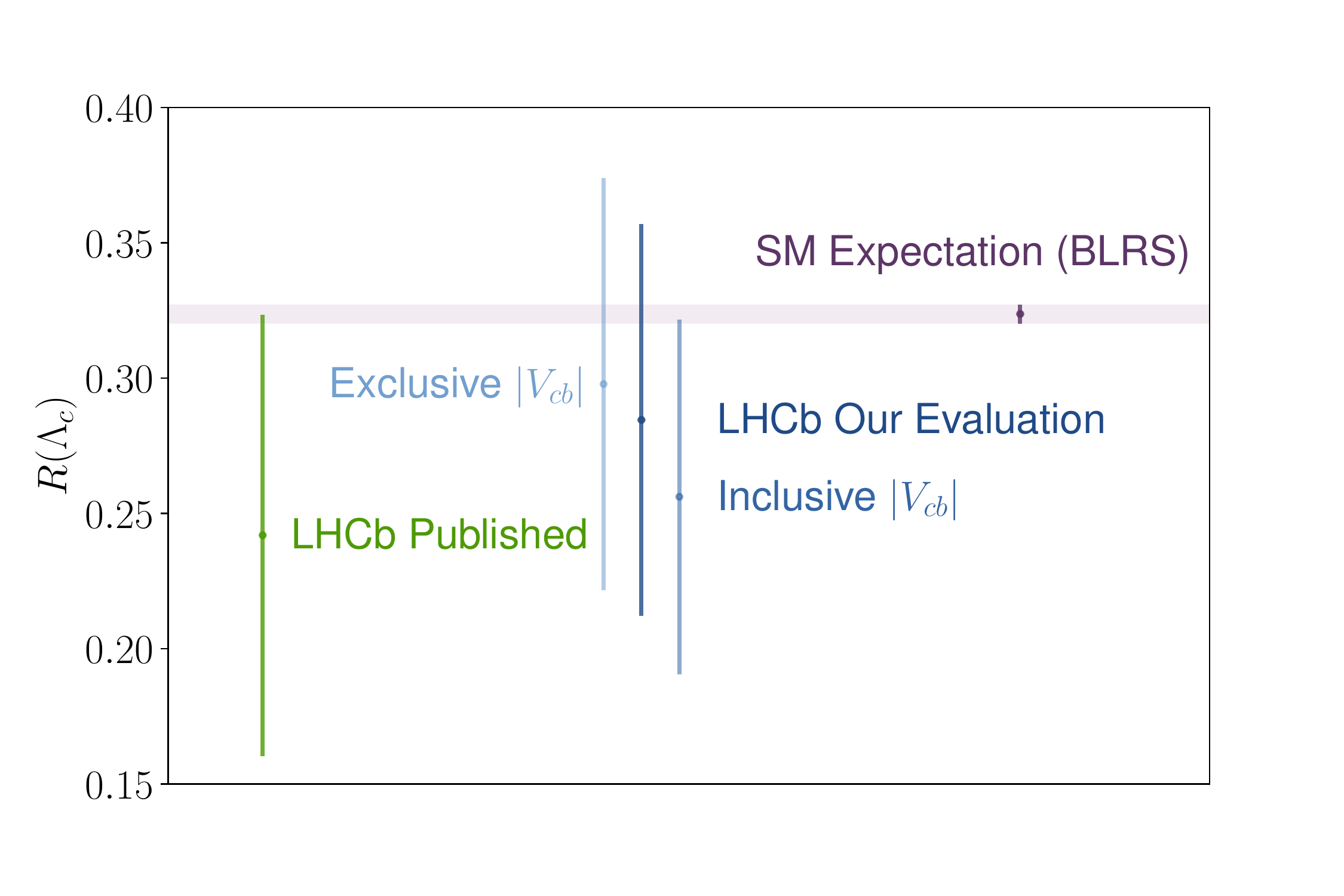}
	\caption{The LHCb~\cite{LHCb:2022piu} result for $R(\Lambda_c)$ (green) is compared to the SM expectation~\cite{Bernlochner:2018kxh, Bernlochner:2018bfn} and to our evaluation (dark blue). We also show $R(\Lambda_c)$ using $|V_{cb}|_{\text{excl}}$ (light blue) and $|V_{cb}|_{\text{incl}}$ (medium blue) from Eq.~(\ref{Vcbval}).
	}
	\label{fig:RLc}
\end{figure}

For tests of LFUV,
one may divide Eq.~\eqref{LHCbtau} by Eq.~\eqref{BLRSpredmu} to obtain (adding uncertainties in quadrature)
\beq\label{RLcNew}
    R(\Lc) = |0.04/V_{cb}|^2\, (0.285 \pm 0.073)\,.
\eeq
This result is in good agreement with the SM prediction in Eq.~\eqref{RLcSM}.  
For comparison, combining the uncertainties in Eq.~\eqref{LHCbresult} in quadrature gives $R(\Lambda_c) = 0.242 \pm 0.076$.
In Fig.~\ref{fig:RLc} we show the various evaluations of $R(\Lambda_c)$,
including the value quoted by LHCb~\cite{LHCb:2022piu} (green),
the SM prediction~\cite{Bernlochner:2018kxh, Bernlochner:2018bfn} (horizontal band), and our updated values for the measured $R(\Lc)$ from Eq.~\eqref{RLcNew},
using $|V_{cb}| = 0.04$ with no uncertainty (dark blue), and 
$|V_{cb}|_{\text{excl}}$ (light blue) and $|V_{cb}|_{\text{excl}}$ (medium blue) from Eq.~(\ref{Vcbval}).

Our result in Eq.~\eqref{RLcNew} has at present only a mildly smaller (absolute or relative) uncertainty, because the last uncertainty in Eq.~\eqref{LHCbtau} is substantial.  A more precise measurement of ${\cal B}(\Lambda_b \to \Lambda_c 3\pi)$ would help reduce the uncertainty. 
The correlation between the SM prediction and our evaluation of $R(\Lambda_c)$ is 6.1\%. 
We also calculate the double ratio to the SM expectation and find $R(\Lambda_c) / R(\Lambda_c)_{\text{SM}} = |0.04/V_{cb}|^2 \, (0.88 \pm 0.22)$,
which can be compared to that obtained from Eq.~\eqref{LHCbresult}, $R(\Lambda_c) / R(\Lambda_c)_{\text{SM}} = 0.75 \pm 0.25$. 

The inclusive semileptonic $\Lambda_b$ widths are predicted to be close to those for $B$ decays~\cite{Manohar:1993qn}.  To ${\cal O}(\lqcd^2/m^2)$ in the OPE, the rates are obtained from the corresponding $B$ decay widths by the replacements of $\lambda_{1,2}$ according to $\lambda_2^{\rm baryon} = 0$ and 
$\lambda_1^{\rm baryon} - \lambda_1^{\rm meson} \simeq 2m_b m_c (m_{\Lambda_b} - \ov{m}_B - m_{\Lambda_c} + \ov{m}_D)/(m_b-m_c) + {\cal O}(\lqcd^3/m^2) \simeq -0.02\,\GeV^2$.
(Here $\ov{m}_B = (3 m_{B^*}+m_B)/4$ is the spin-averaged $B^{(*)}$ mass, and similarly for $\ov{m}_D$.)
With these changes, correcting for the lifetimes, using the isospin averaged measurement ${\cal B}(\Bbar \to X_c\ell\nu) = (10.65 \pm 0.16)\, \%$~\cite{HFLAV:2019otj}, and Ref.~\cite{Ligeti:2014kia} for the $\tau$ mode, we obtain
\begin{subequations}
\begin{align}
    {\cal B}(\Lb \to X_c \mu \bar\nu) & = (10.3 \pm 0.2)\, \%\,,  \label{LbmuSM} \\
    {\cal B}(\Lb \to X_c \tau \bar\nu) & = (2.32 \pm 0.07)\, \%\,. \label{LbtauSM}
\end{align}
\end{subequations}
Comparing the ${\cal B}(\Lambda_b \to \Lambda_c \mu \bar\nu)$ prediction in Eq.~\eqref{BLRSpredmu} to Eq.~(\ref{LbmuSM})
implies that decays to excited states should comprise nearly half of the inclusive rate. 
The prediction in Eq.~(\ref{BLRSpredmu}) is also well below expectations~\cite{Leibovich:2003tw} based on the small-velocity limit.

The CDF measurements of decay rates to the excited states~\cite{Aaltonen:2008eu}
\begin{subequations}
\begin{align}
	\frac{\Gamma[\Lb \to \Lc^*(2595) \mu \bar\nu]}{\Gamma[\Lb \to \Lc \mu \bar\nu]} & = 0.126 \pm 0.033^{+0.047}_{-0.038}\,,\label{eqn:ratioLc12Lc}\\
	\frac{\Gamma[\Lb \to \Lc^*(2625) \mu \bar\nu]}{\Gamma[\Lb \to \Lc \mu \bar\nu]} & = 0.210 \pm 0.042^{+0.071}_{-0.050}\,, \label{eqn:ratioLc32Lc}
\end{align}
\end{subequations}
then appear to imply that further excited states must comprise a surprisingly large fraction of the inclusive rate.
This CDF analysis, however, relies on the isospin limit assumption ${\cal B}(\Lc^*(2595) \to \Lc \pi^+\pi^-) = 2\, {\cal B}(\Lc^*(2595) \to \Lc \pi^0\pi^0)$ to convert the measurement of the reconstructed $\Lc \pi^+\pi^-$ final state to the full branching ratio.
As has been noted in Ref.~\cite{CDF:2011zbc}, the very-near-threshold intermediate resonance $\Lc^*(2595) \to \Sigma_c(2455)\pi$ 
may alter this ratio significantly, such that ${\cal B}(\Lc^*(2595) \to \Lc \pi^+\pi^-) \simeq 0.25\, {\cal B}(\Lc^*(2595) \to \Lc \pi^0\pi^0)$,
although the theory uncertainties in this estimate are not well understood.
This would, however, nominally lead to an enhancement of Eq.~\eqref{eqn:ratioLc12Lc} by a factor of $\simeq 3.3$.
Similar, but far smaller prospective enhancements have also been considered for the $\Lc^*(2625)$ mode~\cite{Nieves:2019nol}, such that the central value in Eq.~\eqref{eqn:ratioLc32Lc} increases to $0.25$.
The smallness of the ${\cal B}(\Lambda_b \to \Lambda_c \mu \bar\nu)$ prediction~\eqref{BLRSpredmu} compared to the inclusive rate prediction suggests that such enhancements may well be present, 
although these particular enhancements,
taken at face value,
would cause the ratio of the  $\Lb \to \Lc^*(2595) \mu \bar\nu$ versus $\Lb \to \Lc^*(2625) \mu \bar\nu$ decays to depart significantly from the (leading order) heavy quark symmetry expectation of $1/2$ 
(see, e.g., Ref.~\cite{Papucci:2021pmj}).
A final state interaction analysis in the $\Lb \to \Lc^* (\to \Sigma_c) \to \Lc$ cascade should be performed to connect HQET predictions to data. 
Further study of such enhancements is therefore well motivated.

In conclusion, we pointed out that normalizing the LHCb measurement of the $\Lambda_b \to \Lambda_c \tau \bar\nu$ rate to the SM prediction for $\Lambda_b \to \Lambda_c \mu \bar\nu$ provides the most robust interpretation for the lepton flavor universality violating ratio, which reduces the significance of a hint for a suppression of $R(\Lambda_c)$.
We presented some evidence that the fraction of excited states in inclusive semileptonic $\Lambda_b$ decay may be significantly greater than in semileptonic $B$ decays, which has important experimental and theoretical implications.

\acknowledgments

We thank Marina Artuso, Marcello Rotondo, Niels Tuning, Guy Wormser, and Wei-Ming Yao for helpful discussions. 
FB is supported by DFG Emmy-Noether Grant No.\ BE~6075/1-1 and BMBF Grant No.\ 05H21PDKBA. 
FB thanks LBNL for its hospitality.
The work of ZL and DJR is supported by the Office of High Energy Physics of the U.S.\ Department of Energy under contract DE-AC02-05CH11231. MP is supported by the U.S.\ Department of Energy, Office of High Energy Physics, under Award Number DE-SC0011632 and by the Walter Burke Institute for Theoretical Physics.

\bibliography{refs}

\end{document}